# Dissociating Cognitive Load and Stress Responses Using Single-Channel EEG: Behavioral and Neural Correlates of Anxiety Across Cognitive States

Neta Batya Maimon [1,2*], Lior Molcho[2], Talya Zaimer [2], Ofir Chibotero [2], Nathan Intrator [1,2], and Eliezer Yahalom [3]

[1]      Tel Aviv University;

[2]      Neurosteer Inc, NYC;

[3]      Brain in Heart TM;

\*      Correspondence: netacoh3@mail.tau.ac.il; Tel.: +972-508816818

Abstract: Identifying neural markers of stress and cognitive load is key to developing scalable tools for mental state assessment. This study evaluated whether a single-channel high-density EEG (hdrEEG) system could dissociate cognitive and stress-related activity during a brief auditory task-based protocol. Sixty-eight healthy adults completed resting-state recordings, cognitively demanding auditory tasks, and exposure to unpredictable lateralized startle stimuli. Participants also rated their stress and anxiety using a modified State–Trait Anxiety Inventory (STAI). EEG analysis focused on frequency bands (Theta, Gamma, Delta) and machine-learning-derived features (A0, ST4, VC9, T2). A double dissociation emerged: Theta and VC9 increased under cognitive load but not startle, supporting their sensitivity to executive function; in contrast, Gamma and A0 were elevated by the startle stimulus, consistent with stress reactivity. ST4 tracked cognitive effort and worry, while T2 negatively correlated with self-reported calmness, indicating relevance to emotional regulation. These results demonstrate that a short, uniform assessment using portable EEG can yield multiple reliable biomarkers of cognitive and affective states. The findings have implications for clinical, occupational, and educational settings, and may inform future neurofeedback protocols targeting simultaneous regulation of attention and stress.

## 1. Introduction

Stress is a physiological and psychological response that occurs when an individual perceives demands that exceed their ability to cope [1]. It can be classified into acute and chronic stress, each with distinct neural and behavioral implications. A closely related concept is anxiety, which is often divided into state anxiety—a temporary emotional response to immediate threats—and trait anxiety, a stable predisposition to experience anxiety across various situations [2]. This distinction is particularly relevant in understanding individual differences in stress reactivity and their impact on cognitive and affective processes.

One of the primary cognitive domains affected by stress is executive function. Executive functions, which include working memory, inhibitory control, and mental flexibility, are crucial for goal-directed behavior [3]. High stress levels have been shown to impair these cognitive processes, reducing performance in attention-demanding tasks and increasing the likelihood of errors [4]. The neurobiological basis of these impairments is well-documented; stress-induced increases in catecholamines such as norepinephrine and dopamine disrupt prefrontal cortex activity, leading to diminished executive control [5]. These effects have been observed both behaviorally and through neuroimaging studies that highlight alterations in prefrontal-limbic connectivity under stress [6].

Researchers have explored various biomarkers to assess stress objectively, particularly those derived from neurophysiological measures such as EEG and fMRI. EEG-based studies have identified stress-related changes in spectral power, with increased gamma and beta-band activity commonly reported in frontal regions during acute stress conditions [7]. Frontal gamma activity has been implicated in emotional arousal and acute stress responses, including increased vigilance and information integration under stress [8-9]. In addition, increased frontal theta activity has consistently been associated with cognitive load, especially during working memory and attention tasks [10-11]. These patterns suggest heightened cortical arousal and reduced inhibition, correlating with self-reported stress levels. Similarly, fMRI studies indicate that acute stress tasks engage the insula, prefrontal cortex, and limbic structures, with stress severity modulating activity in these regions [6]. These neurophysiological markers provide valuable insights into the mechanisms of stress but are rarely assessed alongside cognitive load measures, creating a gap in the literature.

Despite extensive research on cognitive load and stress biomarkers separately, few studies have examined these factors concurrently. Most EEG-based workload assessments focus solely on cognitive task demands without considering the participant's affective state [12]. However, cognitive load and stress often interact in real-world scenarios, affecting performance and neural activity in overlapping ways. For example, both high cognitive load and high stress can increase physiological arousal, complicating the interpretation of EEG signals. The present study addresses this gap by simultaneously measuring cognitive and stress-related biomarkers within a single assessment.

To achieve this, we utilized the Neurosteer high-density resolution EEG (hdrEEG) system, a single-channel device designed for real-time neural monitoring. Previous studies have demonstrated that Neurosteer-derived features, such as A0 and ST4, correlate with cognitive load and cognitive decline. Specifically, A0 has been associated with early cognitive decline and early Parkinson's disease [13-14], while ST4 has been found to correlate with both cognitive decline and individual differences in performance among healthy individuals during working memory tasks [15,13]. Additionally, the system computes novel EEG features, including VC9 and T2. VC9 has been shown to correlate with cognitive load and performance in various experimental paradigms conducted on healthy populations, including numeric and verbal n-back tasks, as well as interruption tasks [15]. It has also been found to distinguish performance in a surgical simulator task among medical interns[15], and to differentiate between resting state and active auditory detection tasks in healthy individuals, patients with cognitive decline, and individuals with Parkinson's disease [13-16]. T2 was extracted from a separate dataset of healthy participants undergoing an auditory mental load task [13] and was found to distinguish between resting state and cognitive load conditions. These biomarkers are derived from advanced signal processing of the single-channel EEG signal, including harmonic analysis and machine learning techniques. While A0 and ST4 have been validated as cognitive state markers, the potential of VC9 and T2 as biomarkers of stress remains underexplored.

In this study, we introduced a startle stimulus within a standard auditory cognitive task to examine whether the extracted single-channel EEG biomarkers differentiate between cognitive effort and stress responses. The startle reflex is a rapid, automatic response to an unexpected sensory event, such as a loud noise, mediated by a subcortical brainstem circuit involving the reticular formation and modulated by higher-order brain regions including the amygdala and prefrontal cortex [17]. In healthy individuals, startle responses are known to activate stress-related neural circuits, leading to transient increases in physiological arousal, heightened alertness, and preparation for action [18]. This response has been linked to increased autonomic activity, including elevated heart rate and skin conductance, and modulations in brain activity, such as enhanced theta and gamma desynchronization in frontal areas, which are measurable through EEG [19]. Importantly, the magnitude of the startle response can vary depending on individual differences in baseline anxiety, attention, and cognitive load, making it a sensitive marker for stress reactivity in non-clinical populations [20].

The goal of this study was to establish whether a single, brief EEG assessment can effectively differentiate cognitive and stress biomarkers. By integrating a startle paradigm and validated self-report measures, we aimed to validate Neurosteer's hdrEEG system as a tool for simultaneous cognitive and stress assessment. If successful, this approach could provide an objective and accessible method for evaluating stress in clinical and occupational settings, where rapid and reliable stress detection is essential.

In the present study, we employed a validated anxiety questionnaire to assess subjective stress levels. The State–Trait Anxiety Inventory (STAI), developed by Spielberger et al. [2], is widely used to measure both transient (state) and enduring (trait) anxiety. The STAI has demonstrated high internal consistency and reliability across various populations, making it a robust tool for self-reported stress evaluation. Integrating such subjective assessments with neurophysiological measures enhances the validity of stress detection methods.

Given the wide variability in how individuals experience and regulate stress, particularly across different age groups, it is essential to develop stress biomarkers that are effective in heterogeneous populations. Our study included a broad adult age range (from 22 to 77) to reflect real-world diversity and assess whether EEG-derived biomarkers can reliably track stress responses across a non-clinical, age-diverse sample. This approach aims to support the development of generalizable tools for stress assessment applicable to varied demographic and occupational settings.

We hypothesized that distinct EEG biomarkers extracted from a single-channel EEG system during a brief, structured assessment could differentially respond to cognitive load and acute stress, reflecting a potential double dissociation between neural signatures of these states. Specifically, we expected that biomarkers such as frontal Gamma activity, as well as ST4 and T2, would be more sensitive to the startle-induced stress condition, while markers like frontal Theta and VC9—previously linked to working memory and cognitive effort—would respond more strongly to cognitively demanding tasks but remain relatively unaffected by the startle stimulus. For exploratory purposes, we also examined how features such as T2 and A0 behave across both stress and cognitive conditions to determine whether one context elicits greater activation. In addition, we assessed whether any of the EEG features correlate with individual differences in stress reactivity and regulation, as measured by subjective reports of perceived stress, control, relaxation ability, and anxiety traits. These measures were derived from the State–Trait Anxiety Inventory (STAI; Spielberger et al.,[2]) and additional validated constructs from the stress and resilience literature. Through this approach, we aimed to characterize the specificity and sensitivity of multiple EEG-derived biomarkers in differentiating between stress and cognitive load states within a short, ecologically valid protocol.

## 2. Materials and Methods

### 2.1. Participants

Ethical approval for the study was granted by the Ethics Committee of Tel Aviv University on May 2, 2024. All participants provided written informed consent prior to participating in the experiment.

Sixty-eight healthy adult volunteers, participated in the study, including 43 women and 25 men. The mean age of the participants was 52.7 years (SD = 14.6). All participants were screened to ensure they had no history of neurological or psychiatric disorders. Informed consent was obtained prior to participation.

2.2. Aparatus

EEG measurements were executed utilizing the Recorder (Neurosteer EEG recorder). An FDA cleared adhesive with three electrodes was applied to the subject's forehead, using a dry gel to enhance signal quality. The non-intrusive electrodes were located at the prefrontal areas, producing a single-EEG channel derived from the difference between Fp1 and Fp2 and a ground electrode at Fpz, based on the international 10/20 electrode positioning. The signal range is ±25 mV (background noise <30nVrms). The electrode contact impedances were kept under 12 kΩ, as determined by a handheld impedance device (EZM4A, Grass Instrument Co., USA). The data was acquired in a continuous mode and subsampled to 500 samples per second.

During the data collection, a proficient research member oversaw each subject to reduce potential muscle interference. Subjects received guidance to refrain from making facial gestures during the session, and the supervising member would notify them if noticeable muscle or eye movements were detected. Notably, the differential signal processing and superior common-mode rejection ratio (CMRR) contribute to minimizing motion disturbances and electrical interference [21].

2.3. Procedure

2.3.1. STAI Questionnaire

Participants first completed a self-report questionnaire based on the State-Trait Anxiety Inventory (STAI). The questionnaire included assessments of state anxiety, reflecting immediate stress levels, and trait anxiety, measuring chronic stress predisposition. Items were rated on a four-point Likert scale, with additional questions assessing momentary stress perception, sleep quality from the previous night, and current emotional state.

2.3.2. EEG recording and auditory assessment protocol

EEG data were recorded using the Neurosteer® single-channel high dynamic range EEG (hdrEEG) Recorder. A three-electrode medical-grade patch was applied to each participant's forehead using dry gel for optimal signal transduction. The monopolar electrode configuration included electrodes positioned at Fp1 and Fp2, based on the International 10/20 system, with a reference electrode at Fpz. The EEG signal was continuously sampled at 500 Hz and transmitted wirelessly for real-time data processing.

EEG signals were processed using time-frequency analysis to extract relevant neural features. Biomarkers A0, ST4, VC0, T2, and VC12 were derived using Neurosteer's proprietary machine-learning algorithms. Previous studies have identified A0 as a marker of cognitive engagement and ST4 as an indicator of cognitive decline and individual task performance, while VC0 and VC12 reflect functional connectivity patterns between EEG frequency bands. The power spectral density of the EEG signal was computed using a fast Fourier transform to analyze delta (0.5–4 Hz), theta (4–7 Hz), alpha (8–15 Hz), beta (16–31 Hz), and gamma (32–45 Hz) frequency bands.

2.3.3. Cognitive Tasks

This study included a previously described auditory detection task (28), an auditory n-back task, and resting state tasks (see Figure 1).

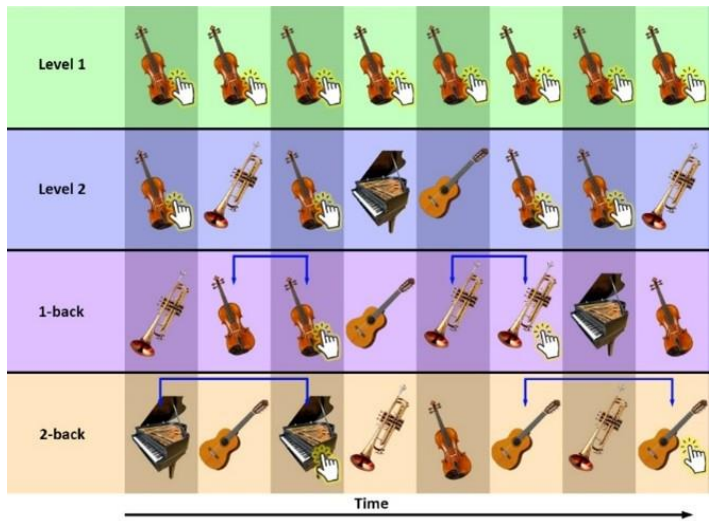

**Figure 1.** A visual representation of the two cognitive tasks used in this study is provided. Auditory Detection: Level 1 features the same melody played by the same musical instrument several times, and the participant is asked to click each time the melody is played. Level 2 presents melodies played by different instruments, and the participant is asked to click only when a melody by a specific instrument is played (in this example, the flute melody). Musical n-back: Levels 1 and 2 showcase melodies played by various instruments. In Level 1, the participant is asked to click whenever a melody is played, while in Level 2, the participant is asked to click only when a melody immediately repeats itself (regardless of which melody is played).

The detection task included a sequence of tunes from a violin, trumpet, and flute. Participants held a clicker in order to respond to the musical cues. Instructions directed participants to click once when they heard a specific instrument playing. Responses were limited to "yes" trials corresponding to the designated instrument's tune. The task was designed with two levels of difficulty to evaluate escalating cognitive demands. In Detection level 1, a consistent tune played for 3 s, recurring throughout the block. Participants were directed to click promptly for every repetition of the tune. This level featured three trials of 90 s each (corresponding to each instrument), with each melody appearing 5–6 times and intervals of 10–18 s of silence. Detection level 2 included tunes lasting for 1.5 s, of three instruments intertwined within a single block. Participants were instructed to respond solely to a designated instrument in that block, disregarding the rest. Each trial in this level had 6–8 melodies interspersed with 8–14 s of silence, and the target tune played 2–3 times.

In the n-back task, participants were presented with a sequence of melodies played by different instruments and used the clicker to respond to the stimuli. This task also included two difficulty levels (0-back and 1-back) to examine increasing cognitive load. A set of melodies (played by a violin, a trumpet, and a flute) was played in a different order for each block, and participants were asked to click a button when the melody repeated n-trials ago (based on the block level). In the 0-back level, participants clicked the button each time a melody was heard. This level included one 90-s block with 9 trials (instances of melody playing), each melody played for 1.5 s and 6–11 s of silence in between. In the 1-back level, participants clicked the button each time a melody repeated itself (n = 1). This level included two 90-s blocks with 12–14 trials (instances of melody playing), each melody played for 1.5 s and 4–6 s of silence in between. In each block, 30–40% of the trials were the target stimulus, where the melody repeated itself, and the participant was expected to click the button.

The resting state condition consisted of three 1-minute trials designed to capture a range of internal states. In the first trial, participants were instructed to keep their eyes open and let their thoughts wander. In the second trial, they were asked to close their eyes and relax, and in the third, they listened passively to a short excerpt of calming meditative music. This design allowed us to record resting neural activity across varying levels of sensory input and cognitive engagement.

To induce acute stress, startle stimuli were presented during one task block per session. These stimuli consisted of lateralized auditory bursts: a ~200 Hz pure sinusoidal tone played for 200 ms with a 50 ms silent gap between bursts. Each burst was delivered either to the left or right ear, with white noise played simultaneously in the opposite ear. The intensity of the tones was approximately 100 dB, and the inter-stimulus interval varied randomly. Acoustic analysis confirmed that both the waveform and frequency spectrum matched these parameters, with a clear peak at ~200 Hz and transient burst patterns (see Figure 2).

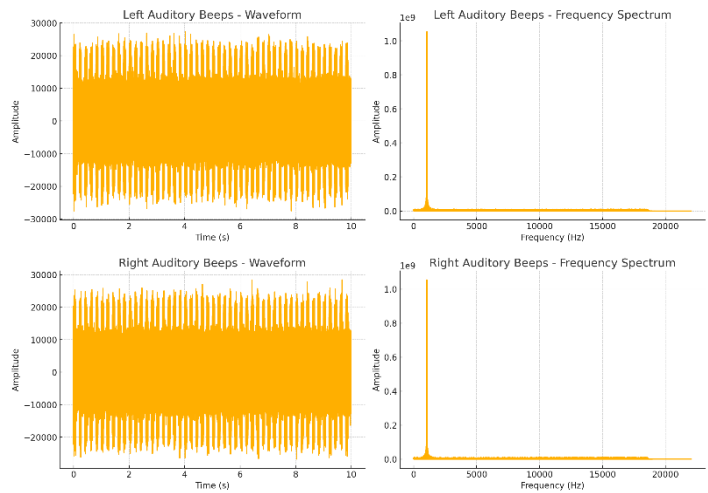

**Figure 2**. Waveform (left) and frequency spectrum (right) of the startle stimuli used in the experiment. The top row displays the left ear condition; the bottom row, the right ear. Each consists of ~200 Hz sine bursts (200 ms duration, 50 ms gap), played unilaterally with simultaneous white noise in the contralateral ear.

2.3.4. Signal Processing

The EEG signal was decomposed into multiple components using harmonic analysis mathematical models [22, 23], and ML methods were employed on the components to extract higher-level EEG features. The full technical specifications for signal processing can be found in Molcho et al. [14]. In summary, the Neurosteer® signal-processing algorithm analyzes EEG data using a time/frequency wavelet-packet analysis. This analysis, previously conducted on a separate dataset of EEG recordings, identified an optimal orthogonal basis decomposition from a large collection of wavelet packet atoms, optimized for that set of recordings using the Best Basis algorithm [24]. This basis results generated a new representation of 121 optimized components called Brain Activity Features (BAFs). Each BAF consists of time-varying fundamental frequencies and their harmonics.

The BAFs are calculated over a 4-s window, which contains 2,048 time elements due to the 500 Hz sampling frequency. In this window, each BAF is a convolution of a time/frequency wavelet packet atom, allowing for a signal that can vary in frequency over the 4-s window, such as a chirp. The window is then advanced by 1 s, similar to a moving window spectrogram with 75% overlap, and the calculation is repeated for the new 4-s window. The EEG power spectrum is obtained using a fast Fourier transform (FFT) of the EEG signals within a 4-s window.

The data was tested for artifacts due to muscle and eye movement of the prefrontal EEG signals (Fp1, Fp2). The standard methods used to remove non-EEG artifacts were all based on different variants of the Independent Components Analysis (ICA) algorithm [25]. These methods could not be performed here, as only a single-channel EEG data was used. As an alternative, strong muscle artifacts have higher amplitudes than regular EEG signals, mainly in the high frequencies; thus, they are clearly observable in many of the BAFs that are tuned to high frequency. This phenomenon helps in the identification of artifacts in the signal. Minor muscle activity is filtered out by the time/frequency nature of the BAFs and thus caused no disturbance to the processed signal. Similarly, eye movements are detected in specific BAFs and are taken into account during signal processing and data analysis.

2.3.5. Construction of Higher-Level Classifiers

Several linear combinations were obtained using ML techniques on labeled datasets previously collected by Neurosteer ® using the described BAFs. Specifically, EEG features VC9 and A0 were calculated using the linear discriminant analysis (LDA) technique [26]. LDA technique is intended to find an optimal linear transformation that maximizes the class separability. LDA models on imaging data were found successful in predicting development of cognitive decline up to 4 years prior to displaying symptoms of decline [27]. EEG feature VC9 was found to separate between low and high difficulty levels of an auditory detection task within healthy participants (ages 20–30). EEG feature A0 was found to separate between resting state with music and auditory detection task within healthy participants (ages 20–30).

EEG feature ST4 was calculated using PCA [28]. Principle component analysis is a method used for feature dimensionality reduction before classification. Studies show that features extracted using PCA show significant correlation to MMSE score and distinguish AD from healthy subjects [29-30], as well as show good performance for the diagnosis of AD using imaging (Choi and Jin, 2018). Here, the fourth principal component was found to separate between low and high difficulty levels of auditory n-back task for healthy participants (ages 30–70). Most importantly, all three EEG features were derived from different datasets than the data analyzed in the present study. Therefore, the same weight matrices that were previously found were used to transform the data obtained in the present study.

The frequency approach has been extensively researched in the past decade, leading to a large body of evidence regarding the association of frequency bands to cognitive functions [31]. In this study, we introduce a novel time-frequency approach for signal analysis and compare it to relevant frequency band results. The EEG features presented here are produced by a secondary layer of ML on top of the BAFs. These BAFs were created as an optimal orthogonal decomposition of time/frequency components following the application of the Best Basis Algorithm [26] on the full wavelet packet tree that was created from a large collection of EEG recordings. Therefore, they are composed of time-varying fundamental frequencies and their harmonics.

As a result of this dynamic nature, and due to the fact that the EEG features are created as linear combinations of multiple BAFs, each feature potentially includes a wide range of frequencies and dynamic varying characteristics. If the time variant characteristic was not present, the spectral envelope of each feature would have represented the full characteristic of each EEG feature. As the time varying component of each BAF is in the millisecond range (sampled at 500 Hz), it is not possible to characterize the dynamics with a spectrogram representation which averages the signal over 4-s windows. Thus, though characterizing the frequency representations of the novel features may be of interest, it is not applicable in this case, much like with EEG-produced ERPs [32-34]. We do observe, however, that EEG feature VC9 includes predominantly fundamental frequencies that belong to the Delta and Theta range (and their harmonics), while EEG features ST4 and A0 are broader combinations of frequencies spanning the whole spectrum (up to 240 Hz).

Other studies conducted on young healthy participants (a different study population than that previously mentioned) showed that EEG feature VC9 activity increased with increasing levels of cognitive load, as manipulated by numeric n-back task [17]. Additionally, VC9 activity during the performance of an arithmetic task decreased with external visual interruptions [35]. VC9 activity was also found to decrease with the repetition of a motor task in a surgery simulator performed by medical interns and was correlated with their individual performance [36]. These studies found that VC9 showed higher sensitivity than Theta, especially for lower-difficulty cognitive loads, which are more suitable for clinical and elderly populations. Within the clinical population, VC9 was found to correlate with auditory mismatch negativity (MMN) ERP component of minimally responsive patients [37]. EEG feature ST4 was found to correlate to individual performance of the numeric n-back task. That is, the difference between high and low load in RTs per participant was correlated to the difference between high and low load in ST4 activity [17].

2.4. Statistical Analysis

EEG data were averaged across trials for each participant and categorized into three task conditions: resting state, startle, and mental load. The resting state condition included all resting state tasks, the startle condition included the abrupt auditory beeps, and the mental load condition combined both the n-back and detection tasks. For each participant, neural activity within these categories was averaged to create a single value per feature and condition.

The primary analysis focused on comparing EEG responses between conditions, evaluating changes in neural biomarkers (A0, ST4, VC0, and T2,) as well as frequency bands (theta, alpha, and beta). Paired t-tests were conducted to assess differences between conditions (e.g., rest vs. mental load). A Benjamini-Hochberg false discovery rate (FDR) correction was applied to control for multiple comparisons.

In addition to these comparisons, Pearson correlations were calculated between EEG biomarker activity and questionnaire responses within each condition. This analysis aimed to identify associations between neural features and subjective reports of stress and anxiety, with a specific focus on composite questionnaire measures such as momentary tension and lack of focus (aggregating relevant items). This approach allowed us to explore how neural markers of cognitive and stress-related activity correspond to subjective experience, both at baseline and following the relaxation intervention.

To control for false positives across multiple statistical tests, correction procedures were applied at several levels. For the paired t-tests comparing EEG features across task conditions, a Benjamini–Hochberg false discovery rate (FDR) correction was implemented, accounting for all EEG features and all pairwise condition comparisons. Only results surviving FDR correction are reported. For correlations between age and questionnaire responses, a Bonferroni correction was applied to the p-values to adjust for the number of questionnaire items tested. In contrast, for the exploratory analysis correlating EEG biomarkers with subjective questionnaire responses, a different criterion was used due to the high dimensionality and exploratory nature of the analysis. Rather than relying solely on p-value thresholds, we reported only those correlations that were consistent across at least two of the three experimental conditions (rest, mental load, startle) for the same EEG feature and questionnaire item. This consistency criterion serves as an indicator of robustness, emphasizing reproducibility across contexts rather than isolated statistical significance. Such an approach has been recommended in exploratory neuroscience studies to mitigate the limitations of conventional correction methods and prioritize replicability [38].

3. **Results**

3.1. Differences between conditions

To examine differences in neural activity between cognitive conditions, we compared EEG biomarkers across the resting state, startle, and mental load conditions. For means and SE of each EEG feature see Figure 3.

For the Theta band, significantly higher activity was observed during mental load compared to rest, $t(1,36) = 6.27$, $p < 0.001$, and significantly lower activity was found during rest compared to startle, $t(1,36) = –4.49$, $p = 0.0002$.

For the VC9 biomarker, activity was significantly higher during mental load compared to rest, $t(1,36) = 6.06$, $p < 0.001$, and significantly lower during rest compared to startle, $t(1,36) = –3.69$, $p = 0.0014$.

For the A0 biomarker, significantly lower activity was observed in the rest condition compared to startle, t(1,36) = –4.85, p = 0.0001, and in the mental load condition compared to startle, t(1,36) = –2.50, p = 0.0317. The comparison between mental load and rest did not reach significance after correction (p = 0.1106).

For the Delta band, activity was significantly higher in the mental load condition compared to rest, t(1,36) = 3.83, p = 0.0010, and significantly lower in rest compared to startle, t(1,36) = –3.10, p = 0.0067.

For the Gamma band, significantly lower activity was observed during rest compared to startle, t(1,36) = –3.98, p = 0.0007.

Finally, for the ST4 biomarker, activity was significantly higher during mental load compared to rest, t(1,36) = 3.53, p = 0.0020. All other comparisons did not reach significance following correction for multiple comparisons.

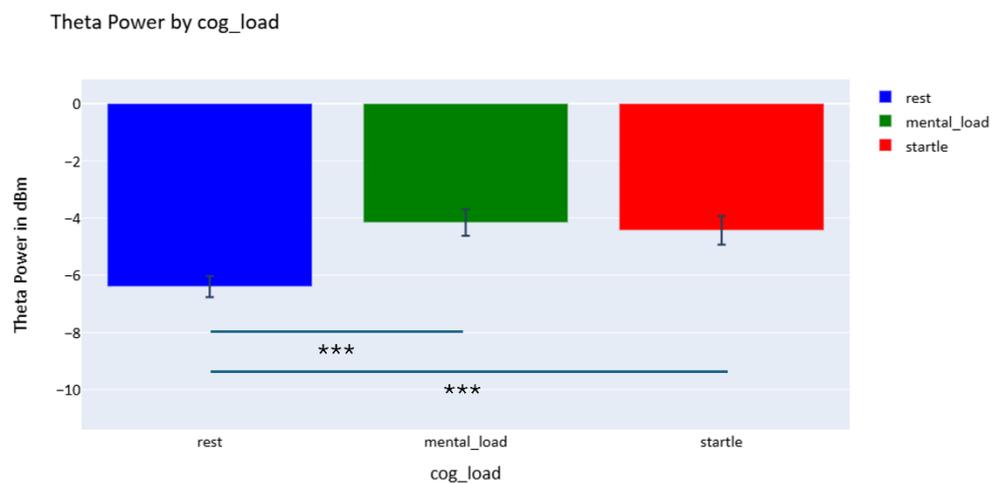

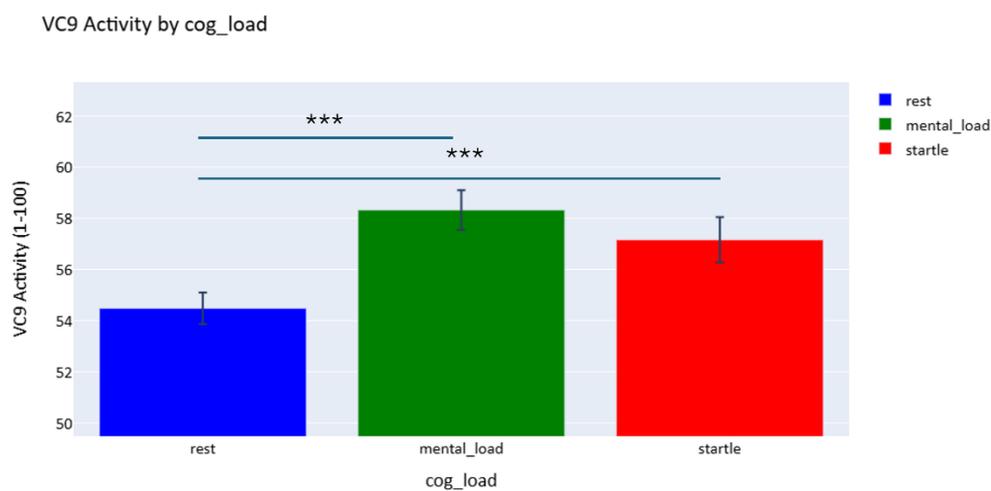

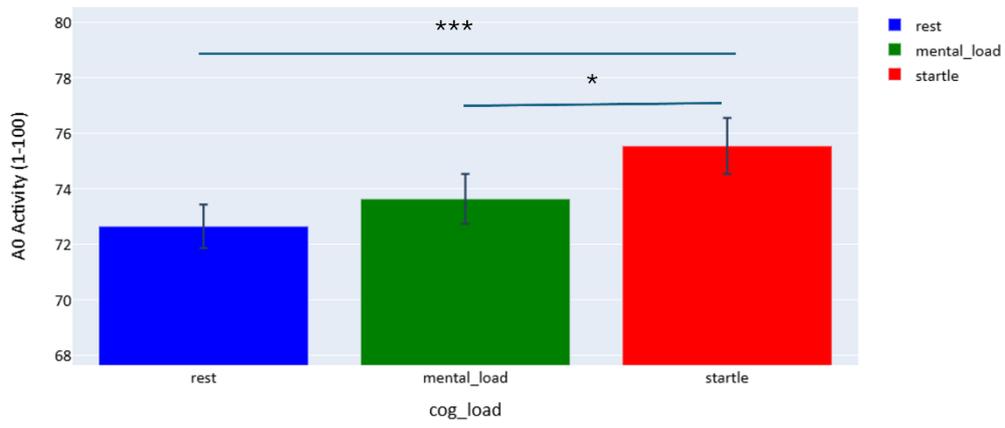

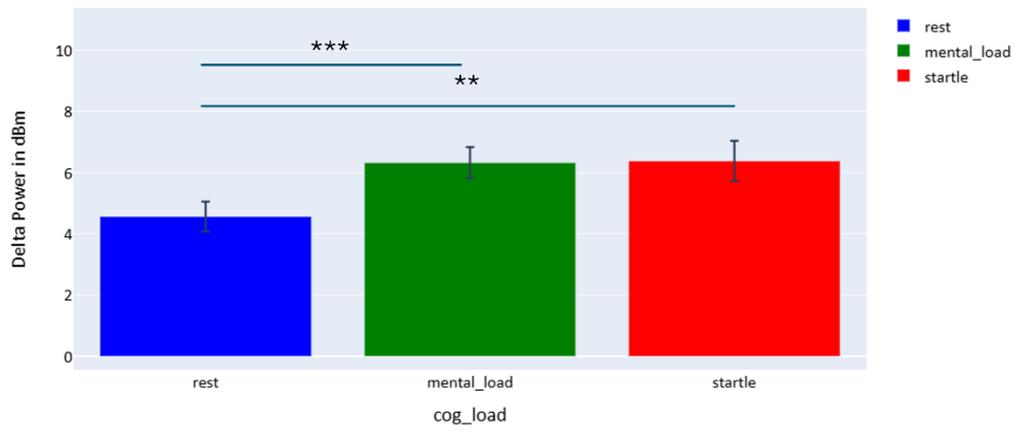

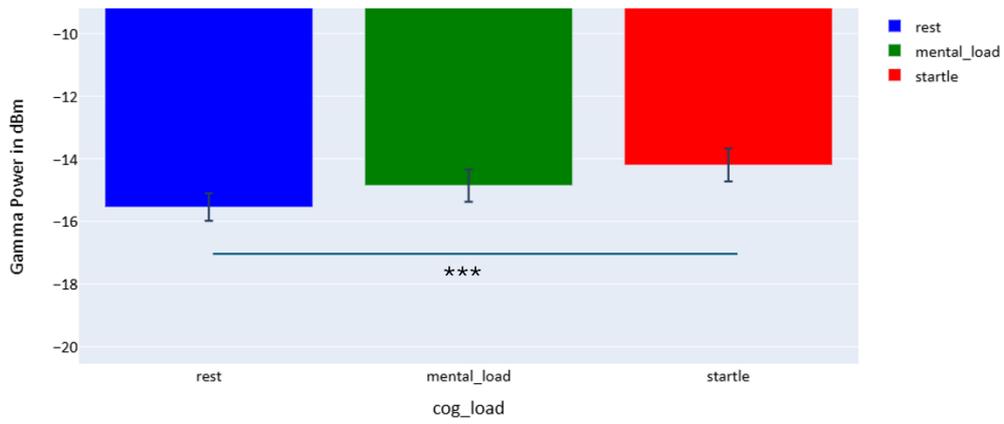

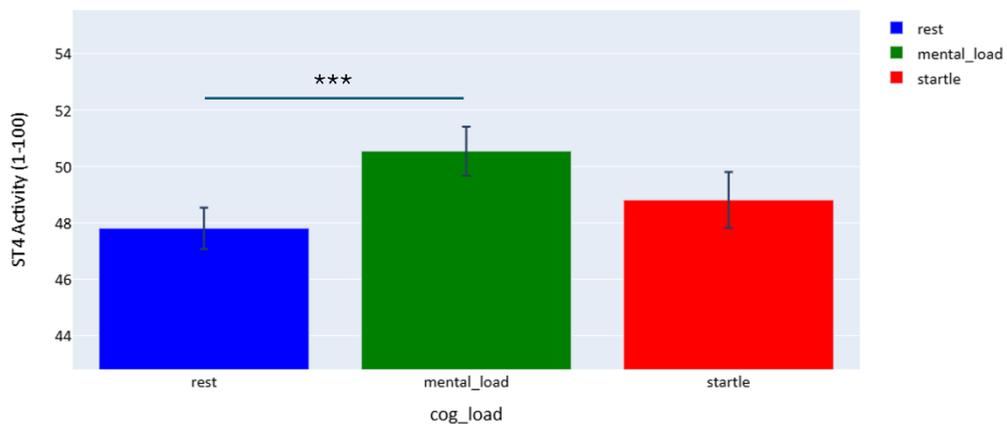

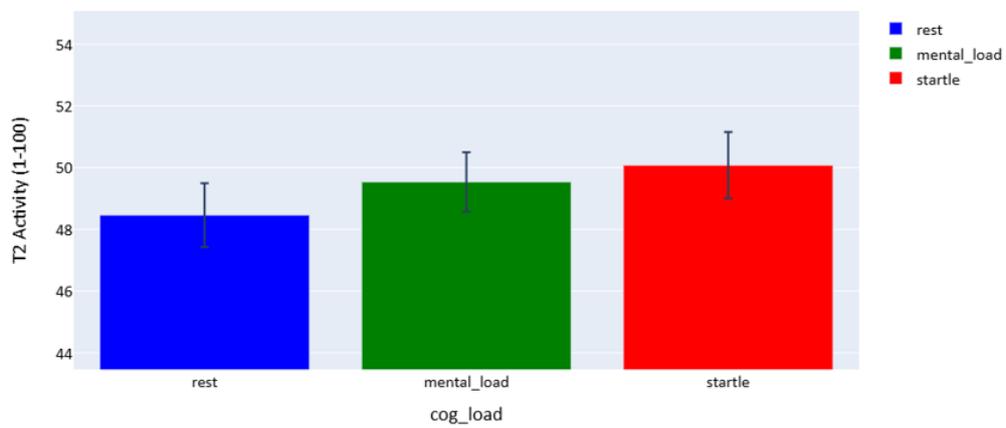

**Figure 3.** Mean activity levels of EEG biomarkers (Theta, VC9, A0, Delta, Gamma, ST4, and T2) across cognitive conditions: rest (blue), mental load (green), and startle (red). Error bars represent standard errors of the mean (SEM). * indicates p < 0.05, ** p < 0.01, *** p < 0.001.

3.2. Questionnaire Responses

To assess individual differences in stress and anxiety, 48 participants completed selected items from the State–Trait Anxiety Inventory (STAI; [2]), including 32 women (M_age = 49.12, SD = 14.00, range: 22–68) and 16 men (M_age = 54.93, SD = 15.38, range: 30–77). Participants also reported their relationship status: 16.9% identified as single, 2.5% as married, 3.6% as divorced, and 3.9% as currently in a relationship.

To assess individual differences in stress and anxiety, participants completed selected items from the State–Trait Anxiety Inventory (STAI; Spielberger et al., [2]). The questionnaire included 10 trait-related items (e.g., "I worry too much about things that don't really matter") and 7 state-related items capturing participants' immediate experience (e.g., "I feel in control right now"). Each item was rated on a 4-point Likert scale, ranging from 1 ("Not at all") to 4 ("Very much").

We calculated the mean and standard deviation for each item to characterize the distribution of responses across participants (table 1).

**Table 1.** Descriptive statistics (mean and standard deviation) for trait and state questionnaire items.

| Index | mean | std |
|---|---|---|
| gen_tens | 2.625 | 0.732963 |
| gen_anx | 1.83333 | 0.833688 |
| gen_wor | 2.40426 | 0.711996 |
| gen_calm | 1.47917 | 0.618495 |
| gen_wor_future | 2.3617 | 0.764015 |
| gen_secure | 3.08333 | 0.70961 |
| gen_hard_tense | 1.97872 | 0.793708 |
| gen_fast_tens | 2.6875 | 0.926128 |
| gen_incon | 1.93617 | 0.818383 |
| gen_relax | 2.85106 | 0.721674 |
| now_tens | 2.10417 | 0.778421 |
| now_wor | 1.93617 | 0.869889 |
| now_calm | 2.83333 | 0.724446 |
| now_restless | 2.06383 | 0.734381 |
| now_no_focus | 2.0625 | 0.782964 |
| now_fear | 1.95745 | 0.779003 |
| now_control | 2.97872 | 0.736896 |
| sleep_hours | 6.2234 | 1.21498 |
| energy_level | 6.7234 | 1.7901 |
| Year Born | 1973.94 | 14.5754 |
| Age | 51.0625 | 14.5754 |

Next, Pearson correlation coefficients were computed between participants' age and each questionnaire item to examine how subjective stress and anxiety relate to age. Two items showed significant correlations with age. The strongest association was observed for the item gen_incon, which reflects general inconsistency in behavior ("I feel inconsistent, I behave differently in similar situations"). This item was negatively correlated with age (r = –0.505, corrected p = 0.0003), indicating that older participants tended to report more stable and consistent behavior patterns. The second strongest correlation was found for the item gen_secure, which reflects a general sense of environmental safety ("I feel safe in my surroundings"). This item was positively correlated with age (r = 0.438, corrected p = 0.0019), suggesting that feelings of safety and groundedness increased with age (see Table 2 for all correlations).

**Table 2.** Pearson correlation coefficients between age and each STAI-based questionnaire item. The table reports the strength and direction of correlations (r) between age and individual questionnaire items, along with the corresponding p-values and FDR-corrected p-values (Benjamini-Hochberg method). Two items reached significance after correction: gen_incon ("I feel inconsistent, I behave differently in similar situations") and gen_secure ("I feel safe in my surroundings"). These findings indicate that increasing age is associated with greater behavioral consistency and perceived environmental safety

| Question | r | p_value | corrected pvalue |
|---|---|---|---|
| gen_incon | -0.505 | 0.0003 | **0.0051** |
| gen_secure | 0.438 | 0.0019 | **0.0323** |
| now_control | 0.384 | 0.0076 | 0.1292 |
| gen_anx | -0.33 | 0.022 | 0.374 |
| gen_wor | -0.327 | 0.0249 | 0.4233 |
| gen_relax | 0.308 | 0.0351 | 0.5967 |
| now_fear | -0.247 | 0.0949 | 1.6133 |
| gen_hard_tense | -0.227 | 0.1255 | 2.1335 |
| now_no_focus | -0.224 | 0.1258 | 2.1386 |
| now_tens | -0.192 | 0.1914 | 3.2538 |
| gen_fast_tens | -0.189 | 0.1977 | 3.3609 |
| gen_wor_future | -0.171 | 0.2499 | 4.2483 |
| gen_tens | -0.123 | 0.404 | 6.868 |
| now_restless | -0.104 | 0.4862 | 8.2654 |
| gen_calm | -0.093 | 0.5292 | 8.9964 |
| now_calm | 0.086 | 0.5628 | 9.5676 |
| now_wor | -0.086 | 0.5645 | 9.5965 |

3.2. Correlations between averaged hdrEEG activity features and subjective ratings

To examine the relationship between EEG biomarkers and subjective reports of stress and cognitive states, Pearson correlations were calculated between participants' condition-averaged neural activity (per EEG feature and condition) and their responses to each question from the State–Trait Anxiety Inventory (STAI), including both state and trait components (10 items total). To ensure robustness, we report only those correlations that were (a) statistically significant (p < 0.05) and (b) consistent across at least two of the three cognitive conditions (rest, mental load, startle) for the same EEG biomarker and questionnaire item. For all significant correlations see Table 3.

**Table 3.** Significant correlations between EEG biomarkers and self-reported questionnaire items across cognitive conditions. This table includes only those correlations that met two criteria: (1) statistical significance at p < 0.05, and (2) consistency across at least two of the three cognitive conditions—rest, mental load, and startle—for the same EEG feature and questionnaire item. EEG biomarkers (ST4, T2, VC9, Delta) were derived from condition-averaged neural activity, and correlated with responses to items from the State–Trait Anxiety Inventory (STAI), including both state and trait components. Yello markers are only present for the robust associations that replicated across conditions.

| task_level | feature | quest | r_value | p_value |
|---|---|---|---|---|
| mental_load | ST4 | now_wor | 0.312448 | 0.043954 |
| mental_load | T2 | gen_relax | -0.47364 | 0.001533 |
| mental_load | VC9 | now_control | -0.33876 | 0.0282 |
| mental_load | Delta | gen_secure | -0.33906 | 0.026135 |
| mental_load | Delta | now_control | -0.34549 | 0.02503 |
| rest | ST4 | now_wor | 0.473556 | 0.001537 |
| rest | T2 | gen_relax | -0.37128 | 0.015488 |
| startle | ST4 | sleep_hours | -0.31006 | 0.04851 |
| startle | T2 | gen_relax | -0.36841 | 0.017786 |
| startle | Delta | gen_relax | -0.37342 | 0.01618 |

T2 activity was negatively correlated with the ability to stay calm in difficult situations (gen_relax), r = –0.47, p = 0.002, indicating that individuals who reported greater calmness under stress showed lower T2 responses during the mental load condition. In the rest condition, a significant negative correlation was also found between T2 activity and general relaxation, r = –0.37, p = 0.015. Similarly, during the startle condition, T2 activity was negatively correlated with this item, r = –0.37, p = 0.018. These results suggest that participants who reported greater relaxation and emotional control in challenging situations exhibited consistently lower T2 activity across all conditions of the auditory assessment (see Figure 4).

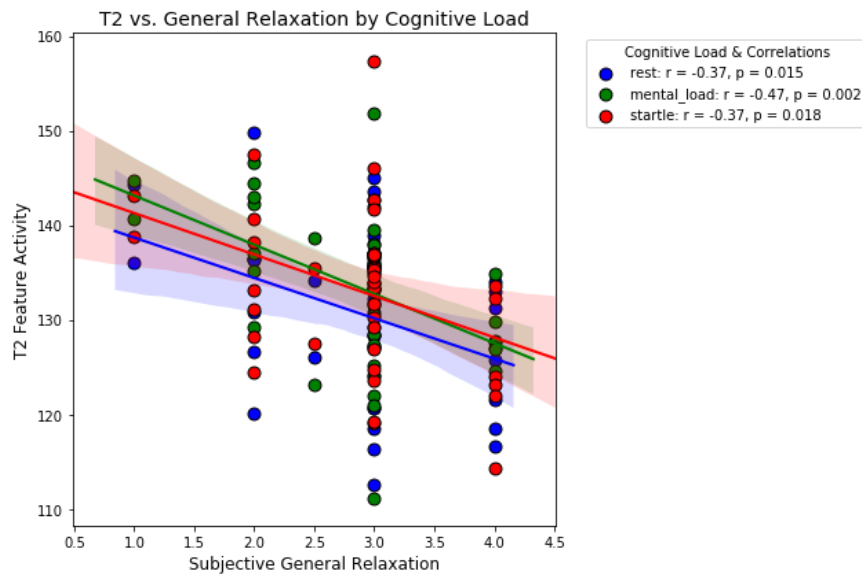

**Figure 4.** Correlation between T2 activity and general relaxation across resting state (blue), mental load (green), and startle (red) conditions. Lower T2 activity was associated with greater self-reported relaxation ability across all conditions.

ST4 activity was positively correlated with momentary worry, r = 0.47, p = 0.002, during the rest condition. A similar positive association was also observed during mental load, r = 0.31, p = 0.044. Although a positive trend was seen in the startle condition, the correlation did not reach statistical significance (r = 0.29, p = 0.066). These findings suggest that individuals reporting higher levels of immediate worry tended to show increased ST4 activity, particularly during rest and mental load conditions (see Figure 5).

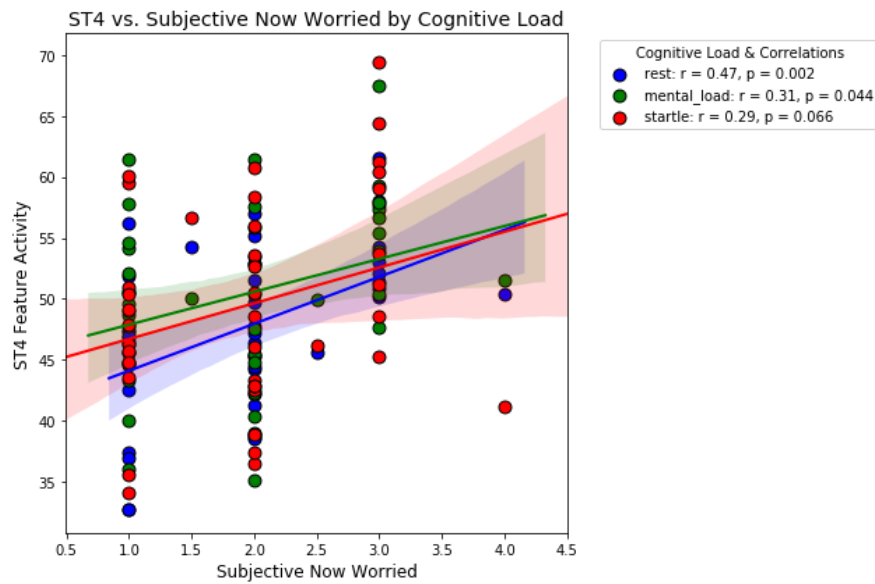

**Figure 5**. Correlation between ST4 activity and momentary worry across resting state (blue), mental load (green), and startle (red) conditions. Lower ST4 activity was associated with greater self-reported momentary worry during mental load and resting state conditions.

## 4. Discussion

The present study examined the capacity of a single-channel high-density resolution EEG (hdrEEG) system to differentiate between neural markers of cognitive load and stress-related activity using a short auditory task-based assessment that included startle stimuli. The results support the potential of this method to detect distinct neural signatures of stress and cognitive effort in a brief and ecologically valid format, with several EEG features showing meaningful relationships to both subjective experiences and experimental manipulations.

Among the most robust findings was the role of Theta band activity, which increased significantly during the mental load condition and decreased during the resting state compared to startle. This aligns with previous literature identifying frontal theta as a neural signature of increased cognitive effort and executive demands. The consistent modulation of theta across both load and stress-related conditions reinforces its relevance in tracking dynamic cognitive engagement.

The VC9 biomarker also demonstrated sensitivity to task-induced mental load. VC9 activity was significantly higher during cognitive load conditions than rest, and positively tracked with other task-relevant features like theta. These findings are consistent with prior research showing VC9's responsiveness to cognitive effort, including in numeric and verbal n-back tasks, surgical simulations, and auditory detection paradigms. The robust differentiation observed in the present study further supports VC9's role as a reliable indicator of cognitive load, particularly in populations where traditional workload measures may be less practical.

Gamma band activity, on the other hand, was uniquely elevated during the startle condition, in line with its proposed association with stress-induced arousal and heightened attentional readiness. Gamma oscillations have been implicated in hypervigilance and acute stress responses, and our results support the use of frontal gamma as a candidate biomarker for rapid stress reactivity.

A0, previously validated as a cognitive state marker, also showed increased activity in the startle condition relative to both rest and cognitive load. This suggests that A0 may capture a combination of arousal and cognitive control processes engaged during high-alert states, such as in response to an unpredictable stimulus. While not specific to cognitive load alone, the elevated A0 during startle aligns with its sensitivity to broad prefrontal activation demands.

ST4, a feature associated with cognitive performance and decline, showed higher activation during mental load compared to rest. In addition to these condition-level effects, ST4 was also positively correlated with participants' momentary worry, particularly during the rest and mental load conditions. This suggests that ST4 may reflect not only general executive demand but also engagement with internal emotional states, including ruminative or anxious thinking.

T2, another task-sensitive feature derived from prior auditory cognitive load paradigms, was uniquely and consistently associated with participants' ability to stay calm in stressful situations. Across all three conditions—rest, mental load, and startle—T2 activity negatively correlated with the self-reported ability to remain composed, indicating that individuals who perceive themselves as more resilient or emotionally regulated tend to exhibit lower T2 responses. This highlights T2's potential as a biomarker for trait-like emotional stability or cognitive-emotional regulation.

Taken together, the results demonstrate a dissociable pattern of neural activity across EEG biomarkers that aligns with our hypothesized differentiation between cognitive load and acute stress responses. Specifically, frontal Theta and VC9 showed consistent increases under cognitive load but were less responsive to startle-induced stress, supporting their role as markers of executive and working memory demands. In contrast, frontal Gamma and A0 activity increased predominantly during the startle condition, suggesting sensitivity to stress-related arousal or heightened attentional readiness. ST4 showed modulation by cognitive demands and was positively associated with subjective worry, particularly during resting and load conditions—pointing to a possible intersection of executive function and emotional engagement. T2, notably, was negatively correlated with the self-reported ability to remain calm across all conditions, suggesting it may reflect trait-like variability in stress reactivity or emotional self-regulation.

These patterns support the feasibility of using a short, task-based protocol with single-channel EEG to capture a range of cognitive and affective processes through a set of distinct, biologically informed biomarkers. The ability to detect double dissociation—where some features respond primarily to cognitive load and others to stress—reinforces the value of using a multi-feature approach. While multichannel EEG or neuroimaging methods provide greater spatial resolution, the current results show that single-channel systems can still yield functionally meaningful insights when combined with targeted cognitive-affective manipulations and robust signal processing. Importantly, these biomarkers may also hold promise for real-time neurofeedback applications. Simultaneous monitoring of stress-related features (e.g., Gamma, A0, T2) alongside cognitive load markers (e.g., Theta, VC9) could enable individualized neurofeedback protocols aimed at enhancing attentional control while reducing stress reactivity. Such dual-target neurofeedback approaches are increasingly supported by empirical research demonstrating improvements in cognitive performance, emotional regulation, and resilience [39]. Future work could explore how dynamic feedback based on these neural markers might support adaptive functioning across a range of applied contexts.

Nonetheless, some limitations remain. Single-channel EEG is inherently more vulnerable to noise, and although preprocessing methods and time-frequency decomposition minimize artifacts, movement or environmental interference may still impact signal quality in uncontrolled settings. Additionally, while our biomarkers are derived using machine learning on independent datasets, interpretation still depends on understanding the participant's context, goals, and emotional state.

Future work should investigate the longitudinal stability of these biomarkers, their predictive power in clinical populations (e.g., PTSD, anxiety disorders), and their responsiveness to therapeutic interventions. Additionally, real-world applications, such as performance monitoring or adaptive cognitive training, could benefit from incorporating EEG features like VC9, T2, or ST4 to personalize support strategies. In conclusion, this study provides initial validation for the use of single-channel hdrEEG in differentiating stress and cognitive load biomarkers. It emphasizes the unique contributions of features like Theta, Gamma, VC9, ST4, A0, and T2, establishing a foundation for accessible, individualized, and scalable brain-based tools for mental performance and well-being.

**Author Contributions**: Conceptualization, E.Y. N.I. and N.M..; methodology, E.Y. and N.M.; software, N.M., T.Z., L.M. and O.C; validation, E.Y.,N.I. and L.M.; formal analysis, N.M.; investigation, N.M.; resources, N.M.; data curation, N.M., T.Z., L.M.; writing—original draft preparation, N.M.; writing—review and editing, E.Y., N.I., N.M., T.Z., L.M. and O.C; visualization, N.M.; supervision, E.Y..; project administration, L.M.; funding acquisition, N.I. All authors have read and agreed to the published version of the manuscript.

**Institutional Review Board Statement**: The study was conducted in accordance with the Declaration of Helsinki and approved by the Institutional Review Board (or Ethics Committee) of Tel Aviv University (protocol code 0002970-2, 2.4.24).

**Informed Consent Statement**: Informed consent was obtained from all subjects involved in the study.

**Conflicts of Interest**: Neta B. Maimon reports equipment was provided by Neurosteer Inc. Neta B. Maimon reports a relationship with Neurosteer Inc that includes: employment. Lior Molcho reports a relationship with Neurosteer Inc that includes: employment. Talya Zeimer reports a relationship with Neurosteer Inc. that includes: employment. Ofir Chibotero reports a relationship with Neurosteer Inc that includes: employment. Nathan Intrator reports a relationship with Neurosteer Inc. that includes: employment and equity or stocks. If there are other authors, they declare that they have no known competing financial interests or personal relationships that could have appeared to influence the work reported in this paper.